\documentclass[superscriptaddress,prl,twocolumn]{revtex4}
\usepackage{amsmath}
\usepackage{amssymb}
\usepackage{amscd}
\usepackage{graphicx,color}
\usepackage[dvipsnames]{xcolor}
\usepackage{bbm}
\usepackage{doi}
\usepackage{dsfont}
\usepackage{datetime}
\usepackage[normalem]{ulem}
\usepackage{bm}

\definecolor{blueberry}{rgb}{0.01569, 0.2, 1.0}
\definecolor{strawberry}{rgb}{1.0,0.0,0.5}

\newcommand{\PO}{\Phi}

\longdate

\begin{document}
\title{Non-reciprocal interactions drive emergent chiral crystallites}

\author{S. J. Kole}
\email{swapnilkole370@gmail.com}
\affiliation{DAMTP, Centre for Mathematical Sciences, University of Cambridge, Wilberforce Rd, CB3 0WA Cambridge, UK}
\affiliation{Isaac Newton Institute for Mathematical Sciences, 20 Clarkson Rd, Cambridge CB3 0EH, UK}

\author{Xichen Chao}
\affiliation{School of Mathematics, University of Bristol, Fry Building, Bristol, BS8 1TW, UK}

\author{Abraham Mauleon-Amieva}
\affiliation{H.H. Wills Physics Laboratory, Tyndall Avenue, Bristol, BS8 1TL, UK}

\author{Ryo Hanai}
\affiliation{Dept. of Physics, Institute of Science Tokyo, 2-12-1 Ookayama Meguro-ku, Tokyo, 152-8551, JAPAN}
\affiliation{Isaac Newton Institute for Mathematical Sciences, 20 Clarkson Rd, Cambridge CB3 0EH, UK}

\author{C. Patrick Royall}
\affiliation{H.H. Wills Physics Laboratory, Tyndall Avenue, Bristol, BS8 1TL, UK}
\affiliation{Gulliver UMR CNRS 7083, ESPCI Paris, Universit´e PSL, 75005 Paris, France}
\affiliation{Isaac Newton Institute for Mathematical Sciences, 20 Clarkson Rd, Cambridge CB3 0EH, UK}

\author{Tanniemola B. Liverpool}
\affiliation{School of Mathematics, University of Bristol, Fry Building, Bristol, BS8 1TW, UK}
\affiliation{Isaac Newton Institute for Mathematical Sciences, 20 Clarkson Rd, Cambridge CB3 0EH, UK}

\begin{abstract}
We study a new type of 2D active material that exhibits  
macroscopic phases with two emergent broken symmetries:
self-propelled {\em achiral} particles that form dense hexatic clusters, 
which {\em spontaneously} rotate. 
We experimentally realise active colloids that self-organise into both polar and hexatic 
crystallites, 
exhibiting exotic emergent phenomena.
This is accompanied by a field theory of coupled order parameters formulated 
on symmetry principles, including non-reciprocity, to capture the non-equilibrium dynamics. We find that the presence of two interacting broken symmetry fields leads to the emergence of novel chiral phases built from (2D) achiral active colloids (here Quincke rollers). These phases are characterised by the presence of both clockwise and counterclockwise rotating clusters. We thus show 
that spontaneous rotation can emerge in non-equilibrium systems, even when the building blocks are achiral, 
due to non-reciprocally coupled broken symmetries. 
This interplay 
leads to self-organized stirring through counter-rotating vortices in confined colloidal systems, with cluster size controlled by external electric fields. 
\end{abstract}

\maketitle

\section{Introduction}

Active matter is a framework to describe a class of complex materials in which the individual components consume energy to perform mechanical work~\cite{marchetti2013hydrodynamics, SRrev}.  These systems show a rich phenomenology, from flocking~\cite{vicsek1995novel,toner1998flocks,toner2005hydrodynamics}, to non-equilibrium phase transitions and novel emergent collective behaviour~\cite{fily2012athermal,cates2015motility,alert2020universal, alert2022active, creppy2015turbulence,dunkel2013fluid,tan2022odd, schwarz2012phase}. While much 
research to date has focused on active fluids with broken rotational symmetries, such as polar and nematic liquid crystal phases, there is increasing recognition that systems with broken translational symmetries or higher-fold orientational order—like hexatic phases—can give rise to exciting and novel active behaviors~\cite{tan2022odd,kole2021layered, kole2024chirality, maitra2019oriented, maitra2020chiral}. Concurrently, a major emerging challenge is {\em{controlling}} activity: selectively activating sub-components of a system, determining the optimal conditions for switching them on or off, and leveraging this control to direct emergent collective behavior toward a desired function, with the ultimate goal of extracting useful work
~\cite{gompper20202020,maggi2017self}. Multiple broken symmetries are a useful tool for this endeavour. 
\textbf{}One promising way is to use {\em active} colloids—microscopically self-driven particles. With active control, the colloids can in principle be designed to self-assemble efficiently ~\cite{bishop2023active,goodrich2017using,zottl2016emergent,sahu2020omnidirectional} paving the way for the design of soft micro-machines ~\cite{zottl2023modeling,aranson2013active,aubret2021metamachines}. 

Nonreciprocity, inherently allowed 
in systems away from equilibrium, is emerging as another pivotal concept in active matter ~\cite{soto2014self,you2020nonreciprocity,saha2020scalar,fruchart2021non, lahiri2000strong}.
Such interactions give rise to exotic phenomena, including odd elasticity and odd viscosity at macroscopic scales \cite{scheibner2020odd, fruchart2023odd, kole2021layered, kole2024chirality}. In driven systems, interactions between different degrees of freedom—such as fields associated  with broken symmetries—are not necessarily symmetric. This non-reciprocal feature drives new non-equilibrium phenemenology, phase transitions and novel pattern forming states~\cite{ frohoff2023non, dinelli2023non, brauns2024nonreciprocal,  suchanek2023entropy}.

\begin{figure*}[hbt!]
\centering
\includegraphics[width=0.95\textwidth]{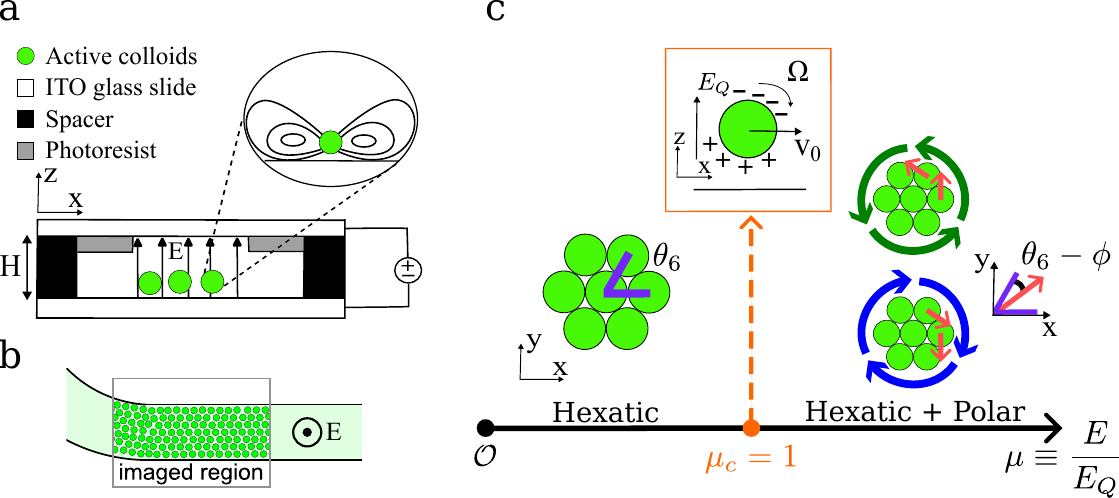}
\caption{Schematic of the self-assembly of active Quincke rollers. (a) Side view of the experimental set up. (b) Top view of experiment showing channel geometry. (c) Plot of the reduced electric field strength, $\mu = E/E_Q$, showing the transition through various regimes. When $E < E_Q$, the colloids organize into a hexatic phase characterized by bond-orientation order ($\theta_6$). At $E = E_Q$, the Quincke instability triggers self-propulsion (polar angle, $\phi$). For $E \geq E_Q$, hexatic clusters begin to rotate in clockwise and counterclockwise directions, with an angular offset between the polar and hexatic bond angles, arising from the non-reciprocal coupling between these two broken symmetries.}
\label{fig1}
\end{figure*}

In this article we present a joint theoretical-experimental study of a 2D active system composed of {\em achiral} self-propelled particles, exhibiting two non-reciprocally coupled order parameters. These order parameters correspond to two broken symmetries: one associated with the rotational symmetry of the particles' self-propulsion direction and the other with the rotational symmetry  of their arrangement around the $\sim 6$ nearest neighbours in the dense phase.
The broken symmetries are characterized by two slow angle variables related to distinct {\em finite} subgroups of $O(2)$: polar angle ($\phi$), and hexatic bond-orientation angle ($\theta_6$), (Fig. \ref{fig1}).  This study presents the first experimental demonstration of a non-reciprocal active colloidal system exhibiting both polar and hexatic order.

We demonstrate the emergence of novel chiral phases, despite being composed of achiral active building blocks (in experiments, \emph{linearly} self-propelled particles). This contrasts with recent work showing macroscopic rotating phases of spinning (chiral) building blocks\cite{tan2022odd,bililign2022motile}. 
Our work reveals that chirality can spontaneously emerge in active systems even when the building blocks are achiral, driven by non-potential interactions that induce rotations at the \emph{mesoscale}. The spontaneous chiral symmetry breaking occurs through non-reciprocal couplings between the two broken symmetries, resulting in the formation of both clockwise and counterclockwise rotating clusters.

An effective 
way to make active colloidal building blocks, is to exploit the Quincke roller mechanism~\cite{bricard2013emergence,geyer2018sounds, chardac2021emergence, PhysRevLett.123.208002, jorge2024active}. Here colloidal 
particles are confined 
to quasi 2D and subjected to a
DC electric field $\vec{E}= E \, \hat{\bf z}$ 
is applied perpendicular to the system, leading to the spontaneous symmetry breaking of the charge distribution at the particle-liquid interface. As a result, rotation at a constant rate emerges from an imposed electric torque acting on the particle (see Fig. \ref{fig1}). For a rigid sphere near to a substrate, the rotation of the particles is coupled with the translation, giving rise to self-propelled {\em achiral} rollers moving in the $xy$-plane (i.e. in 2D), where the speed $v$ is controlled by the strength of the electric field $E=|\vec{E}|$.
We note that although the Quincke mechanism involves rolling, the rotation occurs exclusively in a plane perpendicular to the $xy$-plane, making it irrelevant to the 2D chiral behaviour we study here. Therefore, the trajectories of individual active rollers in the $xy$ plane are {\em not} chiral (i.e. they are isotropic persistent random walks). Conventionally, Quincke rotation is understood to occur above a certain threshold field strength $E_Q$~\cite{mauleon2023dynamics}. Above the critical field strength, $E > E_Q$ single particles undergo continuous rolling leading to active self-propelled motion in two dimensions, Fig. \ref{fig1}. When such particles are studied at high enough concentrations they strongly interact to form a variety of novel collective states such as flocking~\cite{bricard2013emergence,PhysRevLett.123.208002,mauleon2020competing} and motile dense clustered states~\cite{mauleon2020competing}. These motile dense states reveal 
novel non-equilibrium behaviours that are the subject of this paper.

\begin{figure*}[hbt!]
\centering
\includegraphics[width=\textwidth]{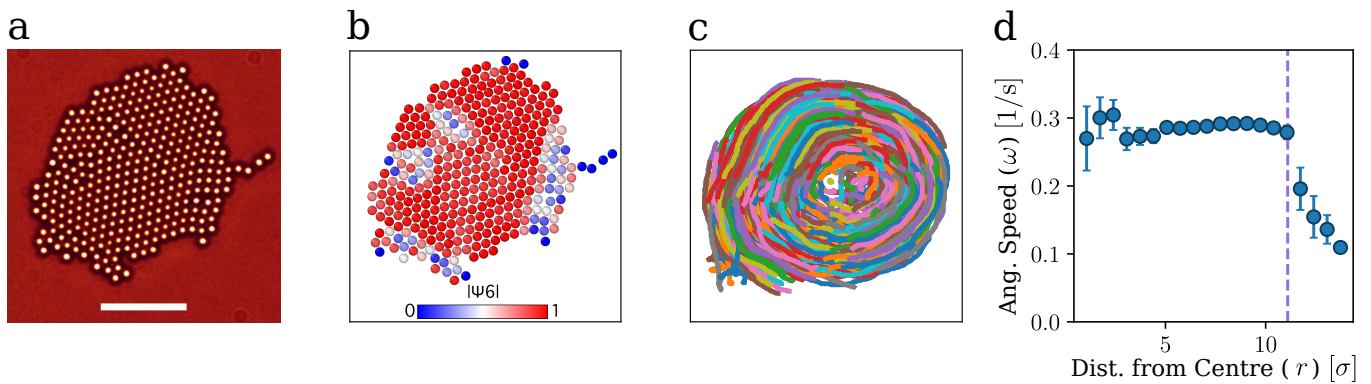}
\caption{Rotating amoeba. (a) Particle resolved image of amoeba cluster, scale bar is 20 $\mu$m. (b) Map of hexatic order parameter visualized in OVITO \cite{stukowski2009visualization}, the neighbour cutoff is the first minimum of the radial distribution function, around 1.55 diameters. (c) Individual particle trajectories within a cluster. (d) Angular velocity as a function of radius is constant showing rigid rotation, dashed line showing the fluid outer boundary of the cluster at around 11 $\sigma$. Error bars give the standard error of the mean (SE)}.
\label{fig2}
\end{figure*}

\section{Experimental System}
\label{sectionExperimental}

We use the so-called Quincke electrorotation mechanism of colloidal active rollers~\cite{bricard2013emergence}. A uniform electric field $E$ aligned in $z-$direction is applied to the suspension. Due to sedimentation, a 2D system forms in the $xy$ plane. 
We use a suspension of colloidal particles of diameter $\sigma=2.92$\ $\mu$m in a non-aqueous ionic solution.

Experiments are performed using 
cells made of two indium-tin-oxide (ITO) coated coverslips 
separated by double-sided tape. The ITO layers are used for the application of the electric field in the $z$-direction.  We use two experimental geometries, which correspond to different density regimes. Our first geometry constitutes a bulk system at low density, i.e. area fraction$\sim0.01$~\cite{mauleon2020competing}. In this case we consider clusters of active rollers which we term ``ameobae''~\cite{mauleon2020competing}.

Secondly, we use a sample cell in the shape of a ``race track'' 
in which we image a section as indicated in Fig. \ref{fig1}(b) which we refer to as a ``channel geometry''. Here the system is a higher density (area fraction $\sim 0.7$). The geometry is defined by the removal of a photoresist layer, as illustrated in Fig.\ref{fig1}.

\textit{Analysis. --- }
We elucidate the local hexagonal order in our crystallites, by using the particle--averaged 
bond orientational order parameter
$\Psi_{6} = \langle \frac{1}{N}\sum_{j}^{N} \psi_{6}^{j} \rangle_{t}$, where $\psi_{6}^j \equiv \frac{1}{z_{j}}\sum_{k=1}^{z_{j}}\exp(i6\theta_{k}^{j})$ quantifies the local order. The quantity $z_{j}$ is the co--ordination number of particle $j$ from a Voronoi tessellation, and $\theta_{k}^{j}$ is the angle between the bond from $i$ to $k$ and a reference axis. The parameter $\psi_{6}^j$ runs between perfect ordering ($\psi_{6} = 1$) and complete disorder ($\psi_{6} =0$). Here the angle brackets denote a time average.

The collective motion of the colloids is characterised by the complex polar order parameter, $\PO = \langle \frac{v_0}{N} \sum_{i}^{N} e^{i \varphi_i}  \rangle_{}$, which quantifies the degree of alignment of the rollers, where  $\mathbf{\hat{u}}_{i} = \left( \cos \varphi_i, \sin \varphi_i \right)$ is the local propulsion orientation of roller $i$ and $v_0$ its speed. $|\PO| = v_0$ indicates perfect azimuthal alignment, and $|\PO| = 0$ represents random particle orientations.

Therefore, the dynamics can be understood as a arising from an interplay of broken rotational symmetries (which give rise to slow modes), namely hexagonal order (indicated by a complex hexatic field $\Psi_6$ with six-fold symmetry) and polar order (indicated by a complex polar field $\Phi$ with vectorial symmetry (see Fig.~\ref{fig1})).

We use two levels of resolution to image our system.
At high resolution, we use particle tracking to obtain the velocities and evaluate the polar order parameter (Fig.~\ref{fig2}). However for lower resolution images, we can obtain better statistics, and here we use particle image velocimetry (PIV) \cite{adrian2011particle} to obtain a velocity field. For the bond orientational order parameter we use particle coordinates throughout. A more extensive description of the experimental setup shown schematically in Fig.\,\ref{fig1}\textbf{a}. is included in  Appendix \ref{sectionExperimentalDetails}.

\section{Theory : emergence of chiral (rotating) states}

In this section, we develop an active field theory for the coupled polar and hexatic order parameters, based on symmetry principles. We demonstrate that introducing terms that explicitly break Onsager symmetry results in the emergence of linearly stable, spontaneously rotating states, characterized by a misalignment between the polar orientation and the hexatic easy axes.


As previously discussed, the two-dimensional dynamics of Quincke roller assemblies can be modeled as a collection of active Brownian particles (ABPs) \cite{bricard2013emergence,mauleon2020competing}, coupled through an attractive electro-hydrodynamic interaction that leads to the formation of hexatic crystallites. In addition, these Quincke rollers exhibit polarised states (flocking).  Consequently, the two relevant order parameters (associated with the broken symmetries) are polarisation and hexatic bond order.
We now develop the field theory for the polarisation field, $\Phi(\bold{x},t)$, and the hexatic ordering field, $\Psi_6(\bold{x},t)\equiv\Psi(\bold{x},t)$ (for brevity), incorporating explicit non-reciprocal couplings. 
We use the complex (Euler) representation for the two fields, which explicitly captures the amplitude and the angle fields as, $\Phi(\bold{x},t)=\Phi_0 e^{i\phi(\bold{x},t)}$ and $\Psi(\bold{x},t)=\Psi_0 e^{i6\theta_6(\bold{x},t)}$. In the ordered phase, the amplitudes quickly relax to constant non-zero values, $\Phi_0$ and $\Psi_0$, while the spatial and temporal dependence are encoded in the polar angle `$\phi$' and hexatic bond angle `$\theta_6$'. The active hydrodynamics of coupled polarisation field, $\Phi(\bold{x},t)$, and hexatic field, $\Psi(\bold{x},t)$, with built-in rotational invariance is

\begin{multline}
      \partial_t \Phi+\lambda_1\left(\Phi\nabla^{*}+\Phi^{*}\nabla\right)\Phi =\\ -\Gamma_{1} \frac{\delta \mathcal{F}}{\delta \Phi^{*}}+\Delta\eta_1 \Phi^{*5}\Psi  +\Delta\eta_3\Phi^{*11}\Psi^2+\xi_{\phi}
      \label{mphieqn}
\end{multline}

\begin{multline}
      \partial_t \Psi +\lambda_2\left(\Phi\nabla^{*}+\Phi^{*}\nabla\right)\Psi =\\ -\Gamma_{2} \frac{\delta \mathcal{F}}{\delta \Psi^{*}} + \Delta\eta_2 \Phi^6 +\Delta\eta_4\Phi^{12}\Psi^{*} +\xi_{\psi}
      \label{mpsieqn}
\end{multline}

    \noindent where $\Gamma_1$ and $\Gamma_2$ are dissipative coefficients, and $\xi_\phi$ and $\xi_\psi$ are complex valued, Gaussian distributed white noises with zero mean and variance given by, $\langle \xi_{\alpha\phi} (\bold{x},t)\xi_{\beta\phi} (\bold{x}',t')\rangle=2\Gamma_2\delta_{\alpha\beta}\delta(\bold{x}-\bold{x}')\delta(t-t')$ and $\langle \xi_{\alpha\psi} (\bold{x},t) \xi_{\beta\psi} (\bold{x}',t')\rangle=2\Gamma_1\delta_{\alpha\beta}\delta(\bold{x}-\bold{x}')\delta(t-t')$ respectively, (indices $\alpha$ and $\beta$ 
     refer to the real and imaginary parts of the respective fields). The terms with coefficients $\lambda_1$ and $\lambda_2$ represent the \emph{active} advection of both the polarisation and the hexatic order by the polarisation field. The asterisk ($*$) denotes the complex conjugate of both ordering fields ($\Phi, \Psi$) and the gradient operator ($\nabla$). In the equilibrium limit, the gradient flow dynamics is completely governed by a free energy functional $\mathcal{F}[\Phi,\Psi]$ (see Appendix \ref{sectiontheory} for the explicit form), with $\Delta\eta_1=\Delta\eta_2=\Delta\eta_3=\Delta\eta_4=0$, and  $\lambda_1=\lambda_2=0$. The terms involving the coefficients $\Delta \eta_i$ can be derived from a free energy description, in which case the coefficients are constrained by Onsager's reciprocal relations. We keep them general to allow for non-reciprocal couplings.

Next, we extract the coupled angle dynamics using the Euler forms of the hexatic and polar order. The temporal derivatives are $ \partial_t\Phi= i\Phi_{0} e^{i\phi}\partial_{t}\phi$ and $ \partial_t\Psi= i6\Psi_{0} e^{i6\theta_6}\partial_{t}\theta_6$. 
The real sectors of Eq. \eqref{mphieqn} and Eq. \eqref{mpsieqn} enforce amplitudes $\Phi_0$ and $\Psi_0$ respectively. While, the imaginary sectors govern the coupled dynamics of the polar angle '$\phi$' and bond angle '$\theta_6$' fields

\begin{multline}
          \partial_{t}\phi +2\lambda_1\Phi_0\left(\cos \phi \partial_x \phi +\sin\phi\partial_y \phi\right) =\Tilde{\eta}_1 \Phi_{0}^{4}\Psi_{0} \sin6(\theta_6-\phi)\\+\Tilde{\eta}_3 \Phi_{0}^{10}\Psi_{0}^2 \sin 12(\theta_6-\phi)+\Gamma_1\sigma_a\nabla^2\phi + \xi'_{\phi}
\end{multline}

\begin{multline}
         \partial_{t}\theta_6 +2\lambda_2\Phi_0\left(\cos\phi \partial_x\theta_6+\sin\phi\partial_y\theta_6\right)=\\\dfrac{1}{6\Psi_{0}}\left[ -\Tilde{\eta}_2 \Phi_{0}^{6} \sin6(\theta_6-\phi)-\Tilde{\eta}_4 \Phi_{0}^{12}\Psi_{0} \sin12(\theta_6-\phi)\right]\\+\Gamma_2\sigma_b\nabla^2\theta_6+\xi'_{\psi} 
\end{multline}

\noindent where $\sigma_a$ and $\sigma_b$ are polar and hexatic elastic contributions respectively and $\Tilde{\eta}_i=\eta_i+\Delta \eta_i$ ($\eta_i$, $\sigma_a$ and $\sigma_b$ originate from the free energy $\mathcal{F}[\Phi,\Psi]$). The redefined noise strengths are $\xi'_\phi=Img[e^{-i\phi}\xi_{\phi}/\Phi_0]$ and $\xi'_\psi=Img[e^{-i6\theta_6}\xi_{\psi}/6\Psi_0]$. When $\Delta \eta_i=0$ (equilibrium condition), it implies the reciprocal constraint $\eta_1/\eta_2=\eta_3/\eta_4=6\Gamma_1/\Gamma_2$. The re-parameterised equations of motion are

\begin{multline}
          \partial_{t}\phi +2\lambda_1\Phi_0\left(\cos \phi \partial_x \phi +\sin\phi\partial_y \phi\right) =\\ \Tilde{\eta}_1 \Phi_{0}^{4}\Psi_{0}\sin6(\theta_6-\phi)+\Tilde{\eta}_3 \Phi_{0}^{10}\Psi_{0}^2 \sin 12(\theta_6-\phi)+\sigma_1\nabla^2\phi +\xi'_{\phi} 
\end{multline}

\begin{multline}
         \partial_{t}\theta_6 +2\lambda_2\Phi_0\left(\cos\phi \partial_x\theta_6+\sin\phi\partial_y\theta_6\right)=\\\dfrac{1}{6\Psi_{0}}\left[-\Tilde{\eta}_2 \Phi_{0}^{6}\sin6(\theta_6-\phi)-\Tilde{\eta}_4 \Phi_{0}^{12}\Psi_{0} \sin12(\theta_6-\phi)\right]\\+\sigma_2\nabla^2\theta_6+\xi'_{\psi} 
\end{multline}

\noindent \noindent where $\sigma_1=\Gamma_1\sigma_a$, $\sigma_2=\Gamma_2\sigma_b$. The Hamiltonian governing the equilibrium dynamics of coupled orientational degrees of freedom is minimised for the zero angle difference ($\theta_6-\phi=0$) \cite{nelson1980solid,drouin2022emergent}. 
For the remainder of the analysis, we retain the mean-field description. It is instructive to analyse the equations of motion in terms of the joint fields $\alpha=\theta_6-\phi$ and $\beta=\theta_6 +\phi$ variables

\begin{multline}
          \partial_{t}\alpha = \left( \dfrac{-\Tilde{\eta}_2 \Phi_{0}^{6}} {6\Psi_0}-\Tilde{\eta}_1 \Phi_{0}^{4}\Psi_{0} \right)\sin6\alpha\\-2\left(\dfrac{\Tilde{\eta}_4 \Phi_{0}^{12}}{6}  +\Tilde{\eta}_3 \Phi_{0}^{10}\Psi_{0}^2 \right) \sin 6\alpha \cos6\alpha 
          \label{gmalpha}
\end{multline}

\begin{multline}
          \partial_{t}\beta = \left( \dfrac{-\Tilde{\eta}_2 \Phi_{0}^{6} }{6\Psi_0}+\Tilde{\eta}_1 \Phi_{0}^{4}\Psi_{0} \right)\sin6\alpha\\+2\left(-\dfrac{\Tilde{\eta}_4 \Phi_{0}^{12}}{6} +\Tilde{\eta}_3 \Phi_{0}^{10}\Psi_{0}^2 \right) \sin 6\alpha \cos6\alpha 
          \label{gmbeta}
\end{multline}

Note from Eq. \ref{gmalpha} and Eq. \ref{gmbeta} that there is only one Nambu-Goldstone mode for the dynamics, which is ($\beta=\theta_6+\phi$), while $\alpha$ is a fast mode that relaxes rapidly. Therefore, the fixed points of the dynamical equations are either $\alpha=0$ (or $\alpha=\pm \pi/6$) with $\partial_{t}\beta=0$ or  

\begin{equation}
    \cos 6 \alpha_0= \dfrac{ -\Tilde{\eta}_2 \Phi_{0}^{6} -6\Tilde{\eta}_1 \Phi_{0}^{4}\Psi_{0}^2 }{2\Psi_0\left(\Tilde{\eta}_4 \Phi_{0}^{12} +6\Tilde{\eta}_3 \Phi_{0}^{10}\Psi_{0}^2 \right)}
    \label{meqfp}
\end{equation}
\noindent  with $\partial_t \beta=\omega_0$

\begin{multline}
    \omega_0= \left( \dfrac{-\Tilde{\eta}_2 \Phi_{0}^{6} }{6\Psi_0}+\Tilde{\eta}_1 \Phi_{0}^{4}\Psi_{0} \right)\sin6\alpha_0\\+2\left(-\dfrac{\Tilde{\eta}_4 \Phi_{0}^{12}}{6} +\Tilde{\eta}_3 \Phi_{0}^{10}\Psi_{0}^2 \right) \sin 6\alpha_0 \cos6\alpha_0
    \label{meqfw}
\end{multline}

A theory of hexatic phases with tilt degrees of freedom \cite{nelson1980solid} corresponds to the equilibrium limit of the coupled orientational order dynamics described above. In the passive case, the tilt angle is slaved to align along a hexatic easy axis, with $\alpha_0=0$ being the only stable solution. Indeed, a rotating state that breaks time reversal symmetry (characterised by a non-zero angle-difference) is only possible in active systems. Notably, Eq. \eqref{meqfp} and Eq. \eqref{meqfw} indicate that both $\pm \alpha_0$  are allowed, permitting both clockwise and counter-clockwise rotation ($\pm \omega_0$). The angular speed ($\omega_0$)is proportional to angle difference and activity. 

We now analyse the linear stability of the rotating state to perturbations, $\alpha=\alpha_0+\delta \alpha$

\begin{multline}
          \partial_{t}(\alpha_{0}+\delta \alpha) = \left( \dfrac{-\Tilde{\eta}_2 \Phi_{0}^{6}} {6\Psi_0}-\Tilde{\eta}_1 \Phi_{0}^{4}\Psi_{0} \right)\sin6(\alpha_{0}+\delta \alpha)\\-2\left(\dfrac{\Tilde{\eta}_4 \Phi_{0}^{12}}{6}  +\Tilde{\eta}_3 \Phi_{0}^{10}\Psi_{0}^2 \right) \sin6 (\alpha_{0}+\delta \alpha)\cos6(\alpha_{0}+\delta \alpha) 
\end{multline}

\noindent which implies the stability condition

\begin{multline}
    6\left( \dfrac{\Tilde{\eta}_2 \Phi_{0}^{6}} {6\Psi_0}+\Tilde{\eta}_1 \Phi_{0}^{4}\Psi_{0} \right)\cos6\alpha_{0}\\+2\left(\dfrac{\Tilde{\eta}_4 \Phi_{0}^{12}}{6}  +\Tilde{\eta}_3 \Phi_{0}^{10}\Psi_{0}^2 \right) (-6+ 12\cos^26\alpha_{0})>0 \;, 
\end{multline}

The active phenomenological coefficients, in principle,  depend on the external electric field. As a result, the stability condition for a specific value of angle difference, $\alpha_0$, is influenced by both the order parameter values and activity. Details of the calculations can be found in the Appendix \ref{sectiontheory}.

Thus, we show that non-reciprocally coupled orientational degrees of freedom lead to a stable finite angle-difference $\alpha_0$, corresponding to a misalignment between the polar order and hexatic easy axes. This results in spontaneous rotation with angular speed $\omega (\alpha_0)$, a phenomenon that can be experimentally tested. For a constant external electric field $E_0$, we predict that clusters identified with constant angle difference will exhibit systematic rigid rotation. The angular speed profile can be experimentally verified, Fig. \ref{fig2}. Furthermore, as rotational degeneracy is spontaneously broken, we expect to observe both clockwise and counter clockwise rotation. To investigate this, we employ a Particle Image Velocimetry (PIV) analysis of the bulk system, (see Fig. \ref{fig3}).

\section{Experiments I: Rotating Amoeba}

Here we consider the experimental system at low area fraction (Sec~\ref{sectionExperimentalDetails}).
In the absence of a field, the particles behave as conventional passive Brownian colloids in $2$D. At low field strengths, while remaining non-motile, particles agglomerate into crystallites with local hexagonal order due to long-ranged attractive interactions which arise from electro-osmotic flows (Fig.\,\ref{fig1}) \cite{yeh1997assembly,ristenpart2004assembly,zhang2004situ}. We emphasize that 
unlike ``living crystals'' \cite{palacci2013living}, and systems exhibiting motility-induced phase separation
\cite{cates2015motility,buttinoni2013dynamical,van2019interrupted}, here the aggregation is driven by long-ranged electrohydrodynamic interactions  \cite{yeh1997assembly,ristenpart2004assembly,zhang2004situ}.

Above a critical field strength $E_{Q}$,  the symmetry of the electric charge distribution at the surface of each colloid breaks spontaneously. As a result, an electric torque acting on the colloids leads to rotation with a constant rate around a random axis transverse to the field $E$ (Fig.\,\ref{fig1}) \cite{quincke1896,pannacci2007rheology}. In the 2D system, rotation couples with translation.  Consequently, 
each colloid becomes an active roller 
exhibiting (achiral) trajectories in the $x-y$ plane. 
As noted in Sec.~\ref{sectionExperimental} above, in the regime $E\gtrsim E_Q$~\cite{quincke1896,pannacci2007rheology,das2013electrohydrodynamic} (Fig.\,\ref{fig1}), we observe that the electro-osmotically generated crystallites transition into a highly mobile active state reminiscent of { `amoebae'}.
These {amoebae}  fascinating dynamic behavior, particularly the theoretically predicted active rotation.  Finally, as we further increase the field strength, these crystallites dissolve into an active gas. Our goal here is to characterise the active behaviour of condensed phases of the active rollers such as the amoeba phase. Within an amoeba, we track long individual particle trajectories and for each long trajectory, measure the angle subtended from the cluster centre and time elapsed, to obtain an angular speed. We are thus able to precisely measure the angular speed as a function of distance from the centre. Our results show that angular speed remains constant within amoeba but decays in the fluidised periphery, where hexatic order is disrupted (see Fig. \ref{fig2}d).

\begin{figure}
\centering
\includegraphics[width=\linewidth]{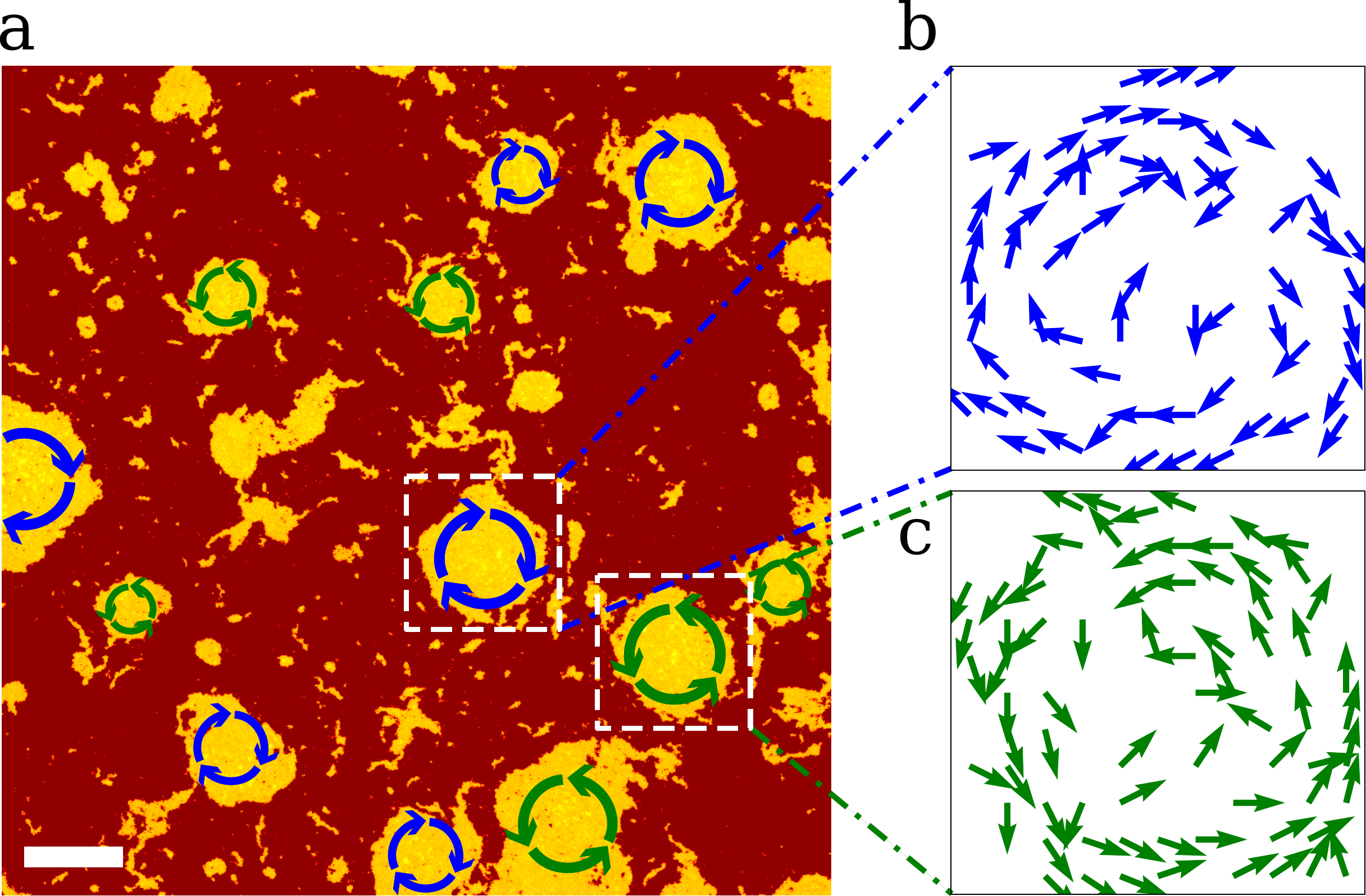}
\caption{Counter rotating amoebae. (a) A snapshot of  the bulk system showing population of rotating clusters. Scale bar is 200 $\mu$m. (b) Velocity field of clockwise rotating cluster generated from PIV (c) Velocity field of cluster rotating anticlockwise generated using  PIV.}
\label{fig3}
\end{figure}

\begin{figure*}[hbt!]
\centering
\includegraphics[width=\linewidth]{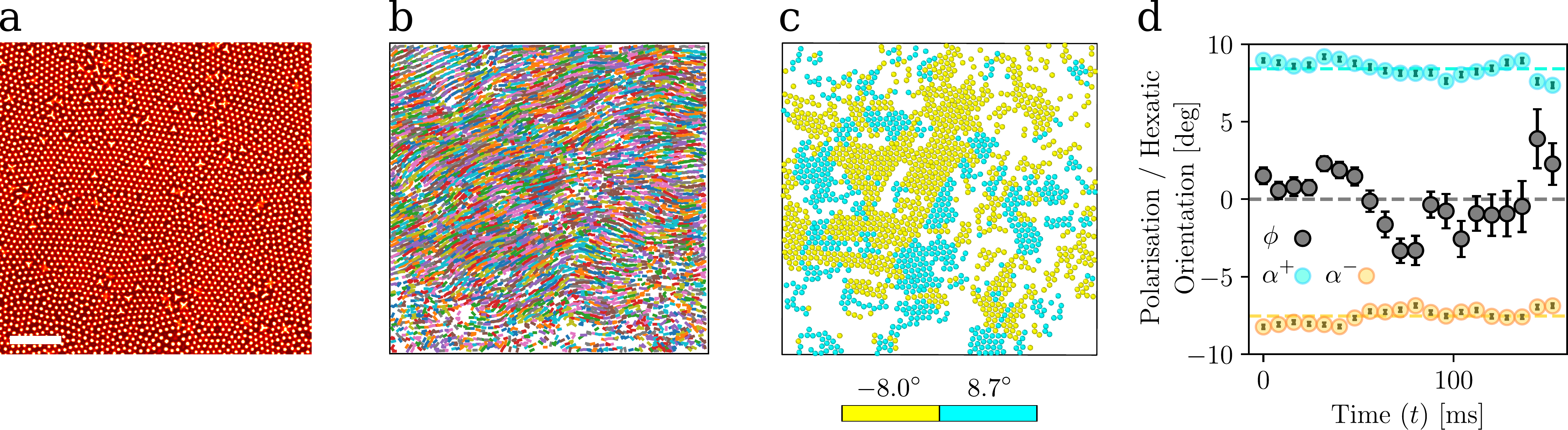}
\caption{Channel Geometry. (a) An experimental snapshot of colloids confined in a channel. Scale bar is 20 $\mu$m. (b) Particle trajectories showing the local polar direction :  the channel edges (which here are aligned roughly with the top and bottom of the panel) guide the colloids thus biasing the polar angle($\phi$). (c) Snapshot showing hexatic domains of fixed $\theta_6$  with orientation relative to the polarisation direction. White regions have low hexatic order/small domains. (d) The relative angle  ($\alpha= \theta_6 - \phi$) 
as a function of time, showing $\theta_6,\phi$.}
\label{fig4}
\end{figure*}

\textit{Rotating amoeba and PIV Analysis. ---}
To further elucidate the underlying flow structures for large system size experiments, we performed Particle Image Velocimetry (PIV) analysis to generate a coarse-grained velocity field of the self-propelled motion of the colloids. Our data reveals the presence of a population of clusters rigidly rotating both clockwise and anticlockwise with associated vortical flows  within each of the clusters, confirming the prediction above (see Fig. \ref{fig3}). These vortical flows emerge as a result of the spontaneous chiral symmetry breaking at the level of individual cluster, allowing for counter-rotating flow profiles. The strength of these flows can be externally controlled with external electric field~\cite{mauleon2020competing,mauleon2023dynamics}.

\section{Experiments II : Channel Geometry}

The amoeba clusters exhibit highly fluctuating polar and hexatic bond orders, 
frequently undergoing coalescence and division events,~\cite{mauleon2020competing}
making it experimentally challenging to quantify the relative angle ($\theta_6-\phi$) in the bulk system. To address this, we conduct 
experiments in which the 
the colloids are confined in a channel geometry at higher density. This configuration enables precise control over the polarization direction, aligning it on average along the channel edges. We see 
a large translating region 
with local hexagonal order with an area fraction that remains constant in time. Confinement in the channel is introduced by patterning the top electrode of the sampling cell. A channel of width $W_{\mathrm{c}} \approx 430 \mu \mbox{m}$ is produced employing conventional photolithography methods. The thickness of the photoresist layer is $\approx 1 \sigma$, sufficient for charge screening on the top electrode. A non--zero electric field, $E$ thus results only within the channel region upon the application of the potential difference.

The electrohydrodynamic attractions and alignment interactions that only occur inside the channel cause the active rollers to condense into a large macroscopic dense homogeneous phase within the channel confinement region. The direction of motion (which sets the angle $\phi$) is controlled by the channel orientation and we can measure the local polarization $\phi$ and measure the hexatic bond angle ($\theta_6$).  We can thus study the angle difference ($\alpha=\theta_6-\phi$) as shown in Fig. \ref{fig4}. Within the observed region, we first group the particles into {\em hexatic domains}. Hexatic domains are defined as regions of high $|\Psi_6|$ with the same bond angle, $\theta_6$ (see Fig. \ref{fig4}c). We find that the hexatic clusters are oriented at angles of $+8.7^\circ$ and $-8.0^\circ$ relative to the polarization direction. It is interesting to note that we find a pair of orientations, $\theta_6^\pm$ consistent with the allowed solutions for the relative angle $\alpha^\pm = \theta_6^\pm - \phi$ (see Eqn. (\ref{meqfp})).

\section{Discussion}

We have presented a joint theoretical-experimental study of an active matter system of interacting Quincke rollers which shows {\em two} competing non-reciprocal $p-$atic orders on  macroscopic scales, namely hexatic bond order and polar (vectorial) order.
We have shown that 
because of non-reciprocal interactions between the two kinds of order, we find the spontaneous formation
of rotating clusters built out of achiral building blocks. This study contrasts with recent studies of  
spinning clusters of chiral fish embryos with hexatic order only~\cite {zhang2010collective}, along with assemblies of spinning colloids in which each spinner constituent already breaks chiral symmetry~\cite{bililign2022motile}.

Our work shows that \emph{chirality can emerge spontaneously} in active systems even if the building blocks are achiral because of rotations emerging at mescoscopic scales. This occurs because of a non-reciprocal coupling between two broken symmetries that can lead to chiral symmetry breaking allowing for both clockwise and anticlockwise rotating clusters.

The velocity field obtained from our PIV measurements shows distinct regions of rotational flow, with counter-rotating vortices, creating a mesoscale self stirring effect. Furthermore, the size of these clusters can be tuned by varying the Peclet number via externally applied electric fields. The size of the colloidal clusters shrinks as the Peclet number increases 
~\cite{mauleon2020competing,mauleon2023dynamics}.

\textit{Perspectives} for the future include the 
incorporation of density fluctuations into our field theory and theoretical and experimental analysis of the transition to the chiral phase. This will involve a study of the microscopic physics controlling the behaviour of the active colloidal particles. To analyse the coalescence and splitting events of clusters will require including the role of long-range hydrodynamic interactions. Finally, the interactions of the rollers with the surface is crucial and needs to be studied in further detail.

\section*{Acknowledgments} The authors would like to thank the Isaac Newton Institute for Mathematical Sciences (INI), Cambridge, for support and hospitality during the programme {\em New statistical physics in living matter}, where part of this work was done. SJK was supported by the INI-Simons research fellowship, Cambridge. The authors thank Michael E. Cates, Sriram Ramaswamy and Ananyo Maitra for valuable discussions. CPR and TBL would like to thank the Kavli Center for Theoretical Physics for hospitality and support during the Active Solids: From Metamaterials to Biological Tissue program. This work was supported by EPSRC grants EP/R014604/1 and EP/T031077/1. XC was supported by CSC-UoB (Bristol) joint PhD Scholarship No. 202108060089. RH was supported by 
was supported by the Grant-in-Aid for Research Activity Start-up from JSPS in Japan (Grant No. 23K19034). CPR would like to acknowledge the Agence National de Recherche for the provision of the grant DiViNew.\\

\paragraph*{Author Contributions:} SJK and TBL conceived the idea. SJK developed and analyzed the hydrodynamic theory. CPR, TBL, SJK, and RH designed the methodology. AMM collected the data, XC analyzed it, and all authors interpreted the results. SJK, TBL, and CPR wrote the manuscript, and all authors edited it.

\paragraph*{Competing interests:} The authors declare that they have no competing interests.

\bibliography{ref}

\maketitle
\onecolumngrid
\appendix
\setcounter{equation}{0}

\section{Experimental Details}
\label{sectionExperimentalDetails}

We use the so-called Quincke electrorotation mechanism of colloidal active rollers~\cite{bricard2013emergence}. A uniform electric field $E$ aligned in the $z-$direction is applied to the suspension. Due to sedimentation, a 2D colloidal system forms in the $xy$ plane. 
We use a suspension of sterically stabilized poly methyl methacrylate
particles of diameter $\sigma=2.92$\ $\mu$m. 
These are suspended in a 0.15-mM solution of dioctyl sulfosuccinate sodium (AOT) in hexadecane. Imaging and application of a uniform dc electric field take place in sample cells made of two indium tin oxide (ITO)- coated glass slides (Solems ITOSOL12), separated with a layer of adhesive tape of thickness 100 $\mu$m. Additionally, a layer of photoresist (Microposit S1818) of 2 $\mu$m in thickness is deposited on the top electrode. In the case of the bulk system, square regions of 5 mm $\times$ 5 mm are created using conventional lithography techniques. In the case of the channel geomentry, a race-track with channels of width 430 $\mu$m was used as shown in Fig.~\ref{fig1}.

The same electric field $E$ that triggers Quincke rotation induces a lateral electric potential gradient between the conductive region and the insulating photoresist layer. As a result, an electrokinetic inward flow confines the rollers at the bottom electrode~\cite{mauleon2020competing}. The electric field is applied by a power supply (Elektro Automatik, PS-2384-05B) and amplified (Trek 606E-6). Image sequences are obtained using brightfield microscopy (Leica DMI 3000B) with a 10× objective and recorded with a frame rate of 354 fps (Basler Ace). All measurements were performed when the rollers reached a steady state.

\textit{Image Analysis --- .}
We use the Trackpy package~\cite{https://doi.org/10.5281/zenodo.1213240, crocker1996methods} to determine particle trajectories, as shown in Fig.~\ref{fig2}(a-c). For the larger sample in Fig.~\ref{fig3}(a), the experimental image resolution is insufficient for particle-level tracking in time, so we use 
Particle Image Velocimetry (PIV)~\cite{adrian2011particle} to construct a coarse-grained velocity field of the colloids' self-propelled motion. A reference frame from the experimental image is chosen, along with another frame captured after a delay. For each small window in the later frame, we identify the most likely displacement and movement direction within a larger window of the reference frame using a least-squares fitting approach.

\section{Theory : emergence of chiral rotating states}

\label{sectiontheory}

\subsection{Active hydrodynamics}

In this section, we formulate and analyse the hydrodynamics of two non-reciprocally coupled order parameters. The two ordering fields are local polarisation $\Phi(\bold{x},t)$ and hexatic order $\Psi(\bold{x},t)$. We use the complex (Euler) representation for the two fields, which explicitly captures the amplitude and the angle modes as, $\Phi(\bold{x},t)=\Phi_0 e^{i\phi(\bold{x},t)}$ and $\Psi(\bold{x},t)=\Psi_0 e^{i6\theta_6(\bold{x},t)}$. In the ordered phase, the amplitudes swiftly relax to constant non-zero values, $\Phi_0$ and $\Psi_0$ and the slow spatial and temporal dependence is encoded in the polar angle `$\phi(\bold{x},t)$' and bond angle `$\theta_6(\bold{x},t)$'. The rotational invariance of angle fields is: $\theta_6\rightarrow \theta_6+\frac{\pi}{3}m$, $m=0, \pm 1...$ and $\phi\rightarrow \phi+2\pi n$, $n=0, \pm 1...$ . Under spatial rotation by an angle $\sigma$, the fields transform as

\begin{equation*}
    \Phi\rightarrow \Phi e^{-i\sigma};
    \Psi\rightarrow \Psi e^{-i6\sigma}\\
    \end{equation*}

\noindent And $\Phi^*\rightarrow \Phi^* e^{i\sigma};
    \Psi^*\rightarrow \Psi^* e^{i6\sigma}$. The equations describing active hydrodynamics of coupled polarisation $\Phi(\bold{x},t)$ and hexatic field $\Psi(\bold{x},t)$ with built-in rotational invariance are

\begin{equation}
      \partial_t \Phi +\lambda_1\left(\Phi\nabla^{*}+\Phi^{*}\nabla\right)\Phi = -\Gamma_{1} \frac{\delta \mathcal{F}}{\delta \Phi^{*}}+\Delta\eta_1 \Phi^{*5}\Psi  +\Delta\eta_3\Phi^{*11}\Psi^2+\xi_{\phi}
      \label{phieqn}
\end{equation}

\begin{equation}
      \partial_t \Psi +\lambda_2\left(\Phi\nabla^{*}+\Phi^{*}\nabla\right)\Psi = -\Gamma_{2} \frac{\delta \mathcal{F}}{\delta \Psi^{*}} + \Delta\eta_2 \Phi^6 +\Delta\eta_4\Phi^{12}\Psi^{*} +\xi_{\psi}
      \label{psieqn}
\end{equation}

\noindent where $\Gamma_1$ and $\Gamma_2$ are dissipative coefficients and $\xi_\phi$ and $\xi_\psi$ are complex valued, Gaussian distributed white noises with zero mean and variance given by, $\langle \xi_{\alpha\phi} (\bold{x},t)\xi_{\beta\phi} (\bold{x}',t')\rangle=2\Gamma_2\delta_{\alpha\beta}\delta(\bold{x}-\bold{x}')\delta(t-t')$ and $\langle \xi_{\alpha\psi} (\bold{x},t)\langle \xi_{\beta\psi} (\bold{x}',t')\rangle=2\Gamma_1\delta_{\alpha\beta}\delta(\bold{x}-\bold{x}')\delta(t-t')$ respectively, (indices $\alpha$ and $\beta$ refer to the real and imaginary parts respectively). The asterisk ($*$) denotes the complex conjugate of both ordering fields ($\Phi, \Psi$) and the gradient operator ($\nabla$). In the equilibrium limit, the dynamics is completely governed by a free energy function $\mathcal{F}[\Phi,\Psi]$. We show below that the 
terms with coefficients $\Delta\eta_1,\Delta\eta_2,\Delta\eta_3,\Delta\eta_4$ can be derived from a free energy description but are restricted to follow Onsager reciprocal relations. While the active case, $\Delta \eta_i$ are unconstrained. Moreover, in the active case, the polar order advects the hexatic field and itself with speeds set by parameters $\lambda_1$ and $\lambda_2$. The most generic active case is $\lambda_1\ne\lambda_2$. The free energy functional $\mathcal{F}[\Phi(\bold{x},t),\Psi(\bold{x},t)]$ is 

\begin{multline}
        \mathcal{F}=\int_{\textbf{x}} a_1|\Phi|^2+b_1|\Phi|^4+c_1|\Phi|^6+d_1|\Phi|^8+e_1|\Phi|^{10}+f_1|\Phi|^{12}\\+\left(a_2+b_2|\Phi|^2+c_2|\Phi|^4+d_2|\Phi|^6+e_2|\Phi|^8+f_2|\Phi|^{10}+g_2|\Phi|^{12}\right)|\Psi|^2+\eta_a\left(\Phi^{*6}\Psi+\Phi^6\Psi^*\right)+\eta_b\left(\Phi^{*12}\Psi^2+\Phi^{12}\Psi^{*2}\right) \\+\sigma_a|\nabla\Phi|^2+\sigma_b|\nabla \Psi|^2
\end{multline}

\noindent where $a_i$, $b_i$, $c_i$, $d_i$, $e_i$, $f_i$, $g_i$, $\eta_a$ and $\eta_b$ are phenomenological coefficients and $\sigma_a$ and $\sigma_b$ are polar and hexatic elastic coefficients respectively. Thus, we obtain the equations of motion 
\begin{multline}
        \partial_t \Phi +\lambda_1\left(\Phi\nabla^{*}+\Phi^{*}\nabla\right)\Phi=\\-\Gamma_1 \bigg(a_{1} +b_{1} |\Phi|^2+ b_{2} |\Psi|^2+c_{1} |\Phi|^4+(c_2+d_{2}|\Phi|^2)|\Psi|^2|\Phi|^2+d_{1} |\Phi|^6+e_{1} |\Phi|^8+e_{2} |\Psi|^2|\Phi|^6+(f_{1} +g_{2}|\Psi|^2|)|\Phi|^{10}+f_{2} |\Psi|^2|\Phi|^8\bigg)\Phi \\+\tilde{\eta}_1 \Phi^{*5}\Psi +\tilde{\eta}_3\Phi^{*11}\Psi^2+\Gamma_1\sigma_a \nabla^2\Phi+\xi_{\phi}
\end{multline}

\begin{multline}
      \partial_t \Psi +\lambda_2\left(\Phi\nabla^{*}+\Phi^{*}\nabla\right)\Psi = -\Gamma_2 \left(a_{2} +b_{2} |\Phi|^2+c_{2} |\Phi|^4+d_{2} |\Phi|^6+e_{2} |\Phi|^8+f_{2} |\Phi|^{10}+g_{2} |\Phi|^{12}\right)\Psi + \tilde{\eta}_2 \Phi^6+\tilde{\eta}_4\Phi^{12}\Psi^{*}\\+\Gamma_2\sigma_b \nabla^2\Psi+\xi_{\psi} 
\end{multline}

\noindent where $\tilde{\eta}_i=\eta_i+\Delta \eta_i$ At equilibrium ($\Delta\eta_1=\Delta\eta_2=\Delta\eta_3=\Delta\eta_4=0$), the Onsager reciprocity implies $\eta_1/\eta_2=\eta_3/\eta_4=6\Gamma_1/\Gamma_2$ (with $\tilde{\eta}_1=\eta_1=-6\Gamma_1\eta_a$, $\tilde{\eta}_2=\eta_2=-\Gamma_2 \eta_a$,  $\tilde{\eta}_3=\eta_3=-12\Gamma_1\eta_b$ and $\tilde{\eta}_4=\eta_4=-2\Gamma_2\eta_b$).  To extract the coupled angle dynamics, we  plug in polar forms of the hexatic and polar order. The temporal derivatives are

\begin{equation}
    \partial_t\Phi= i\Phi_{0} e^{i\phi}\partial_{t}\phi
\end{equation}

\begin{equation}
    \partial_t\Psi= i6\Psi_{0} e^{i6\theta_6}\partial_{t}\theta_6
\end{equation}

 \noindent Assuming spatially uniform $\Phi_0$ and $\Psi_0$, we obtain

\begin{multline}
      i\Phi_{0} e^{i\phi}\partial_{t}\phi +i\Phi_0^{2}\lambda_1\left(e^{i2\phi}\nabla^{*}\phi+\nabla\phi\right) = \left(a_{11} +b_{11} \Phi_{0}^2+ b_{12} \Psi_{0}^2+c_{11} \Phi_{0}^4+c_{12}\Psi_{0}^2 \Phi_{0}^2+d_{11} \Phi_{0}^6+d_{12}\Psi_{0}^2 \Phi_{0}^4\right)\Phi_{0}e^{i\phi}\\+\left(e_{11} \Phi_{0}^8+e_{12}\Psi_{0}^2 \Phi_{0}^6+f_{11} \Phi_{0}^{10}+f_{12}\Psi_{0}^2 \Phi_{0}^8+g_{11} \Phi_{0}^{12}+g_{12}\Psi_{0}^2 \Phi_{0}^{10}\right)\Phi_{0}e^{i\phi}+\tilde{\eta}_1 \Phi_{0}^{5}\Psi_{0}e^{i(6\theta_6-5\phi)} \\+\tilde{\eta}_3\Phi_{0}^{11}\Psi_{0}^2 e^{i(12\theta_6-11\phi)}+\Gamma_1\sigma_a\Phi_0(-e^{i\phi}(\nabla \phi)^2+ie^{i\phi}\nabla^2\phi)+\xi_{\phi}
\end{multline}
\noindent and

\begin{multline}
  6i\Psi_{0} e^{i6\theta_6}\partial_{t}\theta_6 +6i\Phi_0\Psi_0\lambda_2\left(e^{i(6\theta_6+\phi)}\nabla^{*}\theta_6+e^{i(6\theta_6-\phi)}\nabla\theta_6\right) = \\\left(a_{22} + b_{22} \Phi_{0}^2+c_{22} \Phi_{0}^4+d_{22} \Phi_{0}^6+e_{22} \Phi_{0}^8+f_{22} \Phi_{0}^{10}+g_{22} \Phi_{0}^{12}\right)\Psi_{0}e^{i6\theta_6} +\tilde{\eta}_2 \Phi_{0}^{6}e^{i6\phi}\\ +\tilde{\eta}_4 \Phi_{0}^{12}\Psi_{0} e^{i(-6\theta_6+12\phi)}+\Gamma_2\sigma_b\Psi_0(-36e^{i6\theta_6}(\nabla \theta_6)^2+i6e^{i6\theta_6}\nabla^2\theta_6)+\xi_{\psi}
\end{multline}

\noindent Solving for the real and the imaginary part to extract the amplitude and angle field dynamics,

\begin{multline}
      i\Phi_{0} \partial_{t}\phi+i\Phi_0^{2}\lambda_1\left(e^{i\phi}\nabla^{*}\phi+e^{-i\phi}\nabla\phi \right) =\left(a_{11} +b_{11} \Phi_{0}^2+ b_{12} \Psi_{0}^2+c_{11} \Phi_{0}^4+c_{12}\Psi_{0}^2 \Phi_{0}^2+d_{11} \Phi_{0}^6+d_{12}\Psi_{0}^2 \Phi_{0}^4\right)\Phi_{0}\\+\left(e_{11} \Phi_{0}^8+e_{12}\Psi_{0}^2 \Phi_{0}^6+f_{11} \Phi_{0}^{10}+f_{12}\Psi_{0}^2 \Phi_{0}^8+g_{11} \Phi_{0}^{12}+g_{12}\Psi_{0}^2 \Phi_{0}^{10}\right)\Phi_{0} +\tilde{\eta}_1 \Phi_{0}^{5}\Psi_{0}e^{i6(\theta_6-\phi)}\\  +\tilde{\eta}_3 \Phi_{0}^{11}\Psi_{0}^2 e^{i12(\theta_6-\phi)}++\Gamma_1\sigma_a\Phi_0(-(\nabla \phi)^2+i\nabla^2\phi)+\xi_{\phi}e^{-i\phi}
      \label{eqphi}
\end{multline}

\begin{multline}
      6i\Psi_{0} \partial_{t}\theta_6 +6i\Phi_0\Psi_0\lambda_2\left(e^{i\phi}\nabla^{*}\theta_6+e^{-i\phi}\nabla\theta_6\right)\\= \left(a_{22} + b_{22} \Phi_{0}^2+c_{22} \Phi_{0}^4+d_{22} \Phi_{0}^6+e_{22} \Phi_{0}^8+f_{22} \Phi_{0}^{10}+g_{22} \Phi_{0}^{12}\right)\Psi_{0} +\tilde{\eta}_2 \Phi_{0}^{6}e^{i6(\phi-\theta_6)}\\+\tilde{\eta}_4 \Phi_{0}^{12}\Psi_{0} e^{i12(\phi-\theta_6)}+\Gamma_2\sigma_b\Psi_0(-36(\nabla \theta_6)^2+i6\nabla^2\theta_6)+\xi_{\psi}e^{-i6\theta_6}
      \label{eqtheta}
\end{multline}

\noindent Retaining the imaginary part for the coupled polar ($\phi$) and bond angle ($\theta$) dynamics. The real sectors of \eqref{phieqn} and \eqref{psieqn} enforce amplitudes $\Phi_0$ and $\Psi_0$ respectively and the imaginary parts for the coupled polar '$\phi$' and bond angle '$\theta_6$' dynamics, we obtain

\begin{multline}
     \partial_{t}\phi +2\lambda_1\Phi_0\left(\cos \phi \partial_x \phi +\sin\phi\partial_y \phi\right) = \Tilde{\eta}_1 \Phi_{0}^{4}\Psi_{0} \sin6(\theta_6-\phi)+\Tilde{\eta}_3 \Phi_{0}^{10}\Psi_{0}^2 \sin 12(\theta_6-\phi)+\Gamma_1\sigma_a\nabla^2\phi + \xi'_{\phi}
\end{multline}

\begin{multline}
         \partial_{t}\theta_6 +2\lambda_2\Phi_0\left(\cos\phi \partial_x\theta_6+\sin\phi\partial_y\theta_6\right)=\dfrac{1}{6\Psi_{0}}\left[ -\Tilde{\eta}_2 \Phi_{0}^{6} \sin6(\theta_6-\phi)-\Tilde{\eta}_4 \Phi_{0}^{12}\Psi_{0} \sin12(\theta_6-\phi)\right]+\Gamma_2\sigma_b\nabla^2\theta_6+\xi'_{\psi}
\end{multline}

\noindent where $\sigma_a$ and $\sigma_b$ are polar and hexatic elastic contributions respectively and $\Tilde{\eta}_i=\eta_i+\Delta \eta_i$. As discussed before, $\Delta \eta_i=0$ in the equilibrium limit, $\xi'_\phi=Img[e^{-i\phi}\xi_{\phi}/\Phi_0]$ and $\xi'_\psi=Img[e^{-i6\theta_6}\xi_{\psi}/6\Psi_0]$. Note, there is only one Nambu-Goldstone mode for the dynamics. The re-parameterised equations of motion for simplicity

\begin{equation}
          \partial_{t}\phi +2\lambda_1\Phi_0\left(\cos \phi \partial_x \phi +\sin\phi\partial_y \phi\right) = \Tilde{\eta}_1 \Phi_{0}^{4}\Psi_{0}\sin6(\theta_6-\phi)+\Tilde{\eta}_3 \Phi_{0}^{10}\Psi_{0}^2 \sin 12(\theta_6-\phi)+\sigma_1\nabla^2\phi +\xi'_{\phi} 
\end{equation}

\begin{equation}
         \partial_{t}\theta_6 +2\lambda_2\Phi_0\left(\cos\phi \partial_x\theta_6+\sin\phi\partial_y\theta_6\right)=\dfrac{1}{6\Psi_{0}}\left[-\Tilde{\eta}_2 \Phi_{0}^{6}\sin6(\theta_6-\phi)-\Tilde{\eta}_4 \Phi_{0}^{12}\Psi_{0} \sin12(\theta_6-\phi)\right]+\sigma_2\nabla^2\theta_6+\xi'_{\psi} 
\end{equation}

 \noindent where $\sigma_1=\Gamma_1\sigma_a$, $\sigma_2=\Gamma_2\sigma_b$. Now, we move on to analysing the coupled active dynamics of the polar and hexatic bond angle fields. We start by studying the mean-field limit and calculating the fixed points. Its instructive to write the mean-filed dynamics in terms of the joint fields $\alpha=\theta_6-\phi$ and $\beta=\theta_6 +\phi$ as

\begin{equation}
          \partial_{t}\alpha = \left( \dfrac{-\Tilde{\eta}_2 \Phi_{0}^{6}} {6\Psi_0}-\Tilde{\eta}_1 \Phi_{0}^{4}\Psi_{0} \right)\sin6\alpha-2\left(\dfrac{\Tilde{\eta}_4 \Phi_{0}^{12}}{6}  +\Tilde{\eta}_3 \Phi_{0}^{10}\Psi_{0}^2 \right) \sin 6\alpha \cos6\alpha 
\end{equation}

\begin{equation}
          \partial_{t}\beta = \left( \dfrac{-\Tilde{\eta}_2 \Phi_{0}^{6} }{6\Psi_0}+\Tilde{\eta}_1 \Phi_{0}^{4}\Psi_{0} \right)\sin6\alpha+2\left(-\dfrac{\Tilde{\eta}_4 \Phi_{0}^{12}}{6} +\Tilde{\eta}_3 \Phi_{0}^{10}\Psi_{0}^2 \right) \sin 6\alpha \cos6\alpha 
\end{equation}

\noindent The fixed points of the above dynamical equations are either $\alpha=0$ (or $\alpha=\pm \pi/6$) and $\partial_{t}\beta=0$ or  

\begin{equation}
    \cos 6 \alpha_0= \dfrac{ -\Tilde{\eta}_2 \Phi_{0}^{6} -6\Tilde{\eta}_1 \Phi_{0}^{4}\Psi_{0}^2 }{2\Psi_0\left(\Tilde{\eta}_4 \Phi_{0}^{12} +6\Tilde{\eta}_3 \Phi_{0}^{10}\Psi_{0}^2 \right)}
    \label{eqfp}
\end{equation}
\noindent  This implies $\partial_t \beta=\omega_0$

\begin{equation}
    \omega_0= \left( \dfrac{-\Tilde{\eta}_2 \Phi_{0}^{6} }{6\Psi_0}+\Tilde{\eta}_1 \Phi_{0}^{4}\Psi_{0} \right)\sin6\alpha_0+2\left(-\dfrac{\Tilde{\eta}_4 \Phi_{0}^{12}}{6} +\Tilde{\eta}_3 \Phi_{0}^{10}\Psi_{0}^2 \right) \sin 6\alpha_0 \cos6\alpha_0
\end{equation}

Indeed the fixed point in the active case is allowed only when $-\Tilde{\eta}_2 \Phi_{0}^{6} -6\Tilde{\eta}_1 \Phi_{0}^{4}\Psi_{0}^2 < 2\Psi_0\left(\Tilde{\eta}_4 \Phi_{0}^{12} +6\Tilde{\eta}_3 \Phi_{0}^{10}\Psi_{0}^2 \right)$ and $2\Psi_0\left(\Tilde{\eta}_4 \Phi_{0}^{12} +6\Tilde{\eta}_3 \Phi_{0}^{10}\Psi_{0}^2 \right)\neq 0$. The dissipative coefficients $\Gamma_1=\Phi_0^2$, $\Gamma_2=6\Psi_0^2$ together with the Onsager reciprocal relations as discussed before ensure detailed balance.

\subsection{Linear stability analysis of rotating states}

\noindent In this section, we investigate whether the rotating fixed point is linearly stable to perturbations, $\alpha=\alpha_0+\delta \alpha$

\begin{equation}
          \partial_{t}(\alpha_{0}+\delta \alpha) = \left( \dfrac{-\Tilde{\eta}_2 \Phi_{0}^{6}} {6\Psi_0}-\Tilde{\eta}_1 \Phi_{0}^{4}\Psi_{0} \right)\sin6(\alpha_{0}+\delta \alpha)-2\left(\dfrac{\Tilde{\eta}_4 \Phi_{0}^{12}}{6}  +\Tilde{\eta}_3 \Phi_{0}^{10}\Psi_{0}^2 \right) \sin 6(\alpha_{0}+\delta \alpha)\cos6(\alpha_{0}+\delta \alpha) 
\end{equation}

\begin{multline}
    \partial_{t}\delta \alpha = \left( \dfrac{-\Tilde{\eta}_2 \Phi_{0}^{6}} {6\Psi_0}-\Tilde{\eta}_1 \Phi_{0}^{4}\Psi_{0} \right)(\sin6\alpha_{0}\cos6 \delta \alpha+ \cos6\alpha_{0}\sin6\delta\alpha)\\-2\left(\dfrac{\Tilde{\eta}_4 \Phi_{0}^{12}}{6}  +\Tilde{\eta}_3 \Phi_{0}^{10}\Psi_{0}^2 \right) (\sin6\alpha_{0}\cos6 \delta \alpha+ \cos6\alpha_{0}\sin6\delta\alpha)(\cos6\alpha_{0}\cos6 \delta \alpha-\sin6\alpha_{0}\sin6\delta\alpha)
\end{multline}

\noindent Assuming $\delta\alpha$ to be approximately small

\begin{multline}
    \partial_{t}\delta \alpha = \left( \dfrac{-\Tilde{\eta}_2 \Phi_{0}^{6}} {6\Psi_0}-\Tilde{\eta}_1 \Phi_{0}^{4}\Psi_{0} \right)(\sin6\alpha_{0}+ 6\cos6\alpha_{0}\delta\alpha)-2\left(\dfrac{\Tilde{\eta}_4 \Phi_{0}^{12}}{6}  +\Tilde{\eta}_3 \Phi_{0}^{10}\Psi_{0}^2 \right)(\sin6\alpha_{0}+ 6\cos6\alpha_{0}\delta\alpha)(\cos6\alpha_{0}-6\sin6\alpha_{0}\delta\alpha)
\end{multline}

\noindent We now use the fixed point as obtained in \eqref{eqfp} to obtain
\begin{equation}
    \partial_{t}\delta \alpha = \left( \dfrac{-\Tilde{\eta}_2 \Phi_{0}^{6}} {6\Psi_0}-\Tilde{\eta}_1 \Phi_{0}^{4}\Psi_{0} \right)\cos6\alpha_{0}\delta\alpha-2\left(\dfrac{\Tilde{\eta}_4 \Phi_{0}^{12}}{6}  +\Tilde{\eta}_3 \Phi_{0}^{10}\Psi_{0}^2 \right)(-6\sin^{2}6\alpha_{0}\delta\alpha+ 6\cos^26\alpha_{0}\delta\alpha)
\end{equation}

\begin{equation}
    \partial_{t}\delta \alpha = 6\left( \dfrac{-\Tilde{\eta}_2 \Phi_{0}^{6}} {6\Psi_0}-\Tilde{\eta}_1 \Phi_{0}^{4}\Psi_{0} \right)\cos6\alpha_{0}\delta\alpha-2\left(\dfrac{\Tilde{\eta}_4 \Phi_{0}^{12}}{6}  +\Tilde{\eta}_3 \Phi_{0}^{10}\Psi_{0}^2 \right) (-6\delta\alpha + 12\cos^26\alpha_{0}\delta\alpha)
\end{equation}

\noindent The condition for stability is,

\begin{equation}
    6\left( \dfrac{\Tilde{\eta}_2 \Phi_{0}^{6}} {6\Psi_0}+\Tilde{\eta}_1 \Phi_{0}^{4}\Psi_{0} \right)\cos6\alpha_{0}+2\left(\dfrac{\Tilde{\eta}_4 \Phi_{0}^{12}}{6}  +\Tilde{\eta}_3 \Phi_{0}^{10}\Psi_{0}^2 \right) (-6+ 12\cos^26\alpha_{0})>0
\end{equation}

\noindent Let $\Sigma_1=\frac{\Tilde{\eta}_2 \Phi_{0}^{6}} {6\Psi_0}$, $\Sigma_2=\Tilde{\eta}_1 \Phi_{0}^{4}\Psi_{0}$, $\Sigma_3=\Tilde{\eta}_3 \Phi_{0}^{10}\Psi_{0}^2$ and $\Sigma_4=\frac{\Tilde{\eta}_4\Phi_{0}^{12}}{6}$. The inequality condition for a quadratic expression implies two cases

\begin{equation}
    \cos 6\alpha_0>\dfrac{-(\Sigma_2+\Sigma_1)\pm\sqrt{(\Sigma_2+\Sigma_1)^2+32(\Sigma_3+\Sigma_4)^2}}{8(\Sigma_3+\Sigma_4)}
\end{equation}


\end{document}